\documentclass[twoside,leqno,twocolumn]{article}  
\usepackage{ltexpprt}     

\input psfig

\newcommand{\squeezelist}{\setlength{\itemsep}{0pt}}

\def\e{{\varepsilon}}

\def\v1{{v_{i+1}}}
\def\vi2{{v_{i+2}}}
\def\vn1{{v_{n-1}}}
\def\vp1{{v'_{i+1}}}
\def\pie{{\Pi_\varepsilon}}
\def\pixy{{\Pi_{xy}}}

\begin{document}

\title{\Large Locked and Unlocked Polygonal Chains in 
3D\thanks{Reseach supported in part by FCAR, NSERC, and NSF,
and initiated at the Bellairs Research Institute of McGill Univ.
Correspondence to {\tt orourke@cs.smith.edu}.
}
}

\author{
T.\ Biedl\thanks{
        McGill University, Montreal, Canada.
}
\hspace{4mm}
E.\ Demaine\thanks{
        University of Waterloo, Waterloo, Canada.
}
\hspace{4mm}
M.\ Demaine\footnotemark[3]
\hspace{4mm}
S.\ Lazard\footnotemark[2]
\\
A.\ Lubiw\footnotemark[3]
\hspace{4mm}
J.\ O'Rourke\thanks{
        Smith College, Northampton, USA.
}
\hspace{4mm}
M.\ Overmars\thanks{
	Utrecht University, Utrecht, The Netherlands.
}
\hspace{4mm}
S.\ Robbins\footnotemark[2]
\\
I.\ Streinu\footnotemark[4]
\hspace{4mm}
G.\ Toussaint\footnotemark[2]
\hspace{4mm}
S.\ Whitesides\footnotemark[2]
} 

\date{}

\maketitle

\pagestyle{myheadings}
\markboth{}{}


\begin{abstract} \small\baselineskip=9pt 
In this paper, we study movements of simple polygonal chains
in 3D.  We say that an open, simple polygonal chain can be
{\em straightened\/} if it can be continuously
reconfigured to a straight sequence of segments
in such a manner that both the length of each link
and the simplicity of the chain are maintained throughout
the movement.
The analogous concept for closed chains is {\em convexification\/}:
reconfiguration to a planar convex polygon.
Chains that cannot be straightened or convexified are called {\em locked}.
While there are open chains in 3D that are locked, we show that
if an open chain has a simple orthogonal projection
onto some plane, it can be straightened.
For closed chains, we show that there are unknotted but locked closed
chains, and we provide an algorithm for convexifying a planar
simple polygon in 3D with a polynomial number of moves.
\end{abstract}

\section{Introduction}
\label{section:Introduction}
A {\em polygonal chain\/} $P=(v_0,v_1,\ldots,v_n)$ is a sequence
of consecutively joined segments (or edges) 
$e_i =v_iv_{i+1}$ of fixed lengths 
$\ell_i = |e_i|$, 
embedded in space.
A chain is {\em closed\/} (a {\em polygon}) if 
the line segments are joined in cyclic fashion, i.e., if $v_n=v_0$;  
otherwise, it is {\em open}.
Basic questions concerning reconfiguration of open and closed chains
have proved surprisingly difficult.
For example, the question of whether every planar, simple open chain 
can be straightened in the plane while maintaining simplicity 
has circulated in the computational geometry community for years, 
but remains open at this writing.
Previous computational geometry research on the reconfiguration
of chains 
typically concerns planar chains with
crossing links, moving in the presence of obstacles; 
or reconfigures closed chains with crossing links in dimensions $d \ge 2$
\cite{lw-rcpce-95}.
In contrast, throughout this paper we work in 3D 
and
require that chains remain simple throughout their motions.  
The Schwartz-Sharir cell decomposition approach~\cite{ss-pmp2g-83}
from algorithmic robotics
shows that all the problems we consider in this paper are decidable,
and Canny's roadmap algorithm~\cite{c-crmp-87} leads to
solutions that are
singly exponential
in $n$.
Our goal is therefore polynomial-time algorithms.

\section{Open Chains with Simple Projections}
Our first results are algorithms to straighten open polygonal
chains that satisfy either one of two projection conditions.
Our algorithms compute reconfigurations that are 
sequences of ``moves.''
During each move, 
a (small) constant number of individual joint moves occur, 
where for each
a vertex $v_{i+1}$ rotates monotonically 
about an axis through joint $v_i$, 
with the axis of rotation 
fixed in a reference frame attached to some edges.

\vspace{-3pt}
\begin{theorem}
If an open polygonal chain of $n$ links
either has a simple orthogonal projection onto a plane,
or it lies on the surface of a convex polytope,
then it may be straightened in $O(n)$ moves.
The algorithms run in time polynomial in $n$.
\label{theorem:simple.proj}
\end{theorem}

\section{Locked Chains}
\begin{figure}[htbp]
\begin{center}
\ \psfig{figure=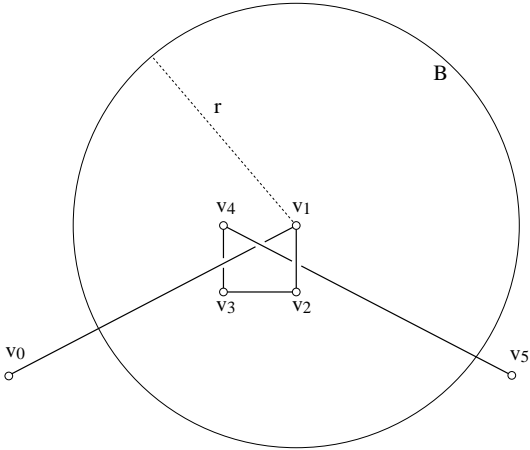,width=7cm}
\end{center}
\caption{A locked open chain $K$ (``knitting needles'').
}
\label{figure:knitting}
\end{figure}

We next show that not all open chains
may be straightened.
Consider the chain $K=(v_0,\ldots,v_5)$ configured as in
Fig.~\ref{figure:knitting}.
One can think of $K$ as
composed of two rigid knitting needles, $e_0$
and $e_4$, connected by a flexible cord of length 
$L =\ell_1+\ell_2+\ell_3$.
By appropriate choice of link lengths and radius $r$ of
a ball $B$ centered on $v_1$,
it can be shown that $v_0$
and $v_5$ remain exterior to $B$ throughout any motion.
This permits completing
a trefoil knot exterior to $B$, which would be unknotted
if $K$ were straightened.  By contradiction, then, $K$ is locked.

By ``doubling'' $K$ and joining endpoints, we prove the same result for
closed chains.
These results were established independently in~\cite{cj-nepiu3s-99}.

\vspace{-3pt}
\begin{theorem}
There exist locked open and locked closed chains.
\end{theorem}

\section{Convexifying Planar Simple Polygons}
A closed chain in a plane, i.e., a planar polygon, may be
convexified in 3D by ``flipping'' out the reflex pockets,
i.e., rotating the pocket chain into 3D and back down to the plane.
This simple procedure was suggested by Erd\H{o}s~\cite{e-p3763-35}
and proved to work by de Sz.~Nagy~\cite{sn-sp3763-39}.
The number of flips, however, cannot be bound as a function
of the number of vertices $n$ of the polygon, as
first proved by Joss and Shannon~\cite{g-hcp-95}.

We offer a new algorithm for convexifying planar closed chains,
which we call the ``St. Louis Arch'' algorithm.
It is more complicated than flipping
but uses a bounded number of moves.  
It models the intuitive approach of picking up the polygon
into 3D.  We discretize this to lifting vertices one by one, accumulating
the attached links into a convex
``arch''
$A$ in a
vertical halfplane above the remaining polygonal chain.  Although the
algorithm is conceptually simple, some care is required to make it
precise, and to then establish that simplicity is maintained
throughout the motions.

Let $P$ be a simple polygon in the $xy$-plane, $\pixy$.
Let $\pie$ be the plane $z = \e$ parallel to $\pixy$, for $\e > 0$.
The value of $\e$ is determined by the initial geometry
of $P$ in a complex way.
We use this plane to convexify the arch safely above the
portion of the polygon not yet picked up.
We use primes to indicate positions of moved (raised) vertices.
Let 
$P[i,j]$ represent the chain 
$(v_i,v_{i+1}, \ldots,v_j)$, including $v_i$ and $v_j$ (where $0\leq i<j<n$),
and let $P(i,j)$ represent the chain without its endpoints.

\begin{figure}[htbp]
\begin{center}
\ \psfig{figure=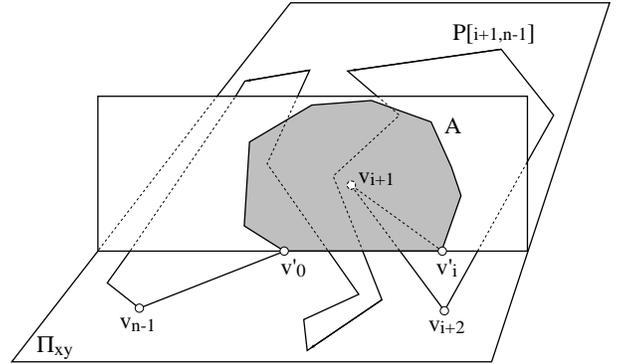,width=8cm}
\end{center}
\caption{The arch $A$ after the $i$th step, i.e.,
after ``picking up'' $P(0,i)$ into $A$.
(The planes $\pixy$ and $\pie$ are not distinguished in
this figure.)
}
\label{figure:A0}
\end{figure}

After a generic step $i$ of the algorithm,
$P(0,i)$ has been lifted above $\pie$ and convexified,
$v_0$ and $v_i$ have been raised to $v'_0$ and $v'_i$ on $\pie$,
and
$P[i+1,n-1]$ remains in its original
position on $\pixy$.
See Fig.~\ref{figure:A0}.

Next $v_{i+1}$ is lifted to $\pie$, the arch $A$ is rotated
down to lie in $\pie$ as well, 
and the resulting ``barbed polygon''
is convexified within $\pie$.
We define a planar polygon as {\em barbed\/} if removal of one ear
leaves a convex polygon, and prove that every barbed polygon
(even ``weakly simple'' ones) 
may be convexified in its plane
in $O(i)$ moves.
After convexification, the arch is rotated up into the vertical
plane containing the new arch base $v'_0 v'_{i+1}$,
and the procedure is repeated.

\begin{theorem}
The ``St.\ Louis Arch'' Algorithm convexifies a planar simple polygon
of $n$ vertices in $O(n^2)$ moves;
it runs in time polynomial in $n$.
\label{theorem:StLouis}
\end{theorem}

\section{Open Problems}
Two of the most prominent among the many open problems suggested by our work
are:
\begin{enumerate}
\squeezelist
\item
What is the complexity of deciding whether a chain (open or closed) 
in 3D is locked?
\item
Can a closed chain with a simple projection always be convexified?
\end{enumerate}

\vspace{-5mm}
\small
\squeezelist
\bibliographystyle{alpha}
\bibliography{lockxxx}

\end{document}